\documentclass[twocolumn]{aastex631}

\begin{document}

\title{Detection of carbon monoxide's 4.6 micron fundamental band structure in WASP-39b's atmosphere with JWST NIRSpec G395H}

\correspondingauthor{David Grant}
\email{david.grant@bristol.ac.uk}

\author[0000-0001-5878-618X]{David Grant}
\affiliation{University of Bristol, HH Wills Physics Laboratory, Tyndall Avenue, Bristol, UK}

\author[0000-0003-3667-8633]{Joshua D. Lothringer}
\affiliation{Department of Physics, Utah Valley University, Orem, UT 84058, USA}

\author[0000-0003-4328-3867]{Hannah R. Wakeford}
\affiliation{University of Bristol, HH Wills Physics Laboratory, Tyndall Avenue, Bristol, UK}

\author[0000-0003-4157-832X]{Munazza K. Alam}
\affiliation{Earth and Planets Laboratory, Carnegie Institution for Science, Washington, DC, USA}

\author[0000-0001-8703-7751]{Lili Alderson}
\affiliation{University of Bristol, HH Wills Physics Laboratory, Tyndall Avenue, Bristol, UK}

\author[0000-0003-4733-6532]{Jacob L. Bean}
\affiliation{Department of Astronomy \& Astrophysics, University of Chicago, Chicago, IL, USA}

\author{Bj\"{o}rn Benneke}
\affiliation{Department of Physics and Institute for Research on Exoplanets, Université de Montréal, Montreal, QC, Canada}

\author[0000-0002-0875-8401]{Jean-Michel D\'esert}
\affiliation{Anton Pannekoek Institute for Astronomy, University of Amsterdam, Amsterdam, The Netherlands}

\author[0000-0002-6939-9211]{Tansu Daylan}
\altaffiliation{LSSTC Catalyst Fellow}
\affiliation{Department of Astrophysical Sciences, Princeton University, Princeton, NJ, USA}

\author[0000-0001-6362-0571]{Laura Flagg}
\affiliation{Department of Astronomy and Carl Sagan Institute, Cornell University, Ithaca, NY, USA}

\author[0000-0003-2215-8485]{Renyu Hu}
\affiliation{Astrophysics Section, Jet Propulsion Laboratory, California Institute of Technology, Pasadena, CA, USA}
\affiliation{Division of Geological and Planetary Sciences, California Institute of Technology, Pasadena, CA, USA}

\author[0000-0001-9164-7966]{Julie Inglis}
\affiliation{Division of Geological and Planetary Sciences, California Institute of Technology, Pasadena, CA, USA}

\author[0000-0002-4207-6615]{James Kirk}
\affiliation{Department of Physics, Imperial College London, London, UK}

\author[0000-0003-0514-1147]{Laura Kreidberg}
\affiliation{Max Planck Institute for Astronomy, Heidelberg, Germany}

\author[0000-0003-3204-8183]{Mercedes L\'opez-Morales}
\affiliation{Center for Astrophysics $\vert$ Harvard \& Smithsonian, Cambridge, MA, USA}

\author[0000-0002-9428-8732]{Luigi Mancini}
\affiliation{Department of Physics, University of Rome "Tor Vergata", Rome, Italy}
\affiliation{INAF - Turin Astrophysical Observatory, Pino Torinese, Italy}
\affiliation{Max Planck Institute for Astronomy, Heidelberg, Germany}

\author[0000-0001-5442-1300]{Thomas Mikal-Evans}
\affiliation{Max Planck Institute for Astronomy, Heidelberg, Germany}

\author[0000-0002-0502-0428]{Karan Molaverdikhani}
\affiliation{University Observatory Munich, Ludwig Maximilian University, Munich, Germany}
\affiliation{Exzellenzcluster Origins, Garching, Germany}

\author[0000-0003-0987-1593]{Enric Palle}
\affiliation{Instituto de Astrof\'isica de Canarias (IAC), 38200 La Laguna, Tenerife, Spain}
\affiliation{Deptartamento de Astrof\'isica, Universidad de La Laguna (ULL), 38206 La Laguna, Tenerife, Spain}

\author[0000-0002-3627-1676]{Benjamin V.\ Rackham}
\altaffiliation{51 Pegasi b Fellow}
\affiliation{Department of Earth, Atmospheric and Planetary Sciences, Massachusetts Institute of Technology, Cambridge, MA 02139, USA}
\affiliation{Kavli Institute for Astrophysics and Space Research, Massachusetts Institute of Technology, Cambridge, MA 02139, USA}

\author[0000-0003-3786-3486]{Seth Redfield}
\affiliation{Astronomy Department and Van Vleck Observatory, Wesleyan University, Middletown, CT, USA}

\author[0000-0002-7352-7941]{Kevin B. Stevenson}
\affiliation{Johns Hopkins APL, Laurel, MD, USA}

\author{Jeff Valenti}
\affiliation{Space Telescope Science Institute, Baltimore, MD, USA}

\author[0000-0003-0354-0187]{Nicole L. Wallack}
\affiliation{Earth and Planets Laboratory, Carnegie Institution for Science, Washington, DC, USA}

\author[0000-0002-7004-8670]{Keshav Aggarwal}
\affiliation{Indian Institute of Technology Indore, India}

\author[0000-0003-0973-8426]{Eva-Maria Ahrer}
\affiliation{Centre for Exoplanets and Habitability, University of Warwick, Coventry, UK}
\affiliation{Department of Physics, University of Warwick, Coventry, UK}

\author{Ian J.M. Crossfield}
\affiliation{Department of Physics and Astronomy, University of Kansas, Lawrence, KS, USA}

\author[0000-0001-7866-8738]{Nicolas Crouzet}
\affiliation{Leiden Observatory, University of Leiden, Leiden, The Netherlands}

\author[0000-0003-2329-418X]{Nicolas Iro}
\affiliation{Institute for Astrophysics, University of Vienna, Vienna, Austria}

\author[0000-0002-6500-3574]{Nikolay K. Nikolov}
\affiliation{Space Telescope Science Institute, Baltimore, MD, USA}

\author[0000-0003-1452-2240]{Peter J. Wheatley}
\affiliation{Centre for Exoplanets and Habitability, University of Warwick, Coventry, UK}
\affiliation{Department of Physics, University of Warwick, Coventry, UK}

\collaboration{31}{(JWST Transiting Exoplanet Community ERS team)}

%% Note that the \and command from previous versions of AASTeX is now
%% depreciated in this version as it is no longer necessary. AASTeX 
%% automatically takes care of all commas and "and"s between authors names.

%% AASTeX 6.31 has the new \collaboration and \nocollaboration commands to
%% provide the collaboration status of a group of authors. These commands 
%% can be used either before or after the list of corresponding authors. The
%% argument for \collaboration is the collaboration identifier. Authors are
%% encouraged to surround collaboration identifiers with ()s. The 
%% \nocollaboration command takes no argument and exists to indicate that
%% the nearby authors are not part of surrounding collaborations.

%% Mark off the abstract in the ``abstract'' environment. 
\begin{abstract}
Carbon monoxide (CO) is predicted to be the dominant carbon-bearing molecule in giant planet atmospheres, and, along with water, is important for discerning the oxygen and therefore carbon-to-oxygen ratio of these planets. The fundamental absorption mode of CO has a broad double-branched structure composed of many individual absorption lines from 4.3 to 5.1 \textmu m, which can now be spectroscopically measured with JWST. Here we present a technique for detecting the rotational sub-band structure of CO at medium resolution with the NIRSpec G395H instrument. We use a single transit observation of the hot Jupiter WASP-39b from the JWST Transiting Exoplanet Community Early Release Science (JTEC ERS) program at the native resolution of the instrument ($R \,{\sim} 2700$) to resolve the CO absorption structure. We robustly detect absorption by CO, with an increase in transit depth of 264\,$\pm$\,68\,ppm, in agreement with the predicted CO contribution from the best-fit model at low resolution. This detection confirms our theoretical expectations that CO is the dominant carbon-bearing molecule in WASP-39b's atmosphere, and further supports the conclusions of low C/O and super-solar metallicities presented in the JTEC ERS papers for WASP-39b. 
\end{abstract}

%% Keywords should appear after the \end{abstract} command. 
%% The AAS Journals now uses Unified Astronomy Thesaurus concepts:
%% https://astrothesaurus.org
%% You will be asked to selected these concepts during the submission process
%% but this old "keyword" functionality is maintained in case authors want
%% to include these concepts in their preprints.

%% From the front matter, we move on to the body of the paper.
%% Sections are demarcated by \section and \subsection, respectively.
%% Observe the use of the LaTeX \label
%% command after the \subsection to give a symbolic KEY to the
%% subsection for cross-referencing in a \ref command.
%% You can use LaTeX's \ref and \label commands to keep track of
%% cross-references to sections, equations, tables, and figures.
%% That way, if you change the order of any elements, LaTeX will
%% automatically renumber them.
%%
%% We recommend that authors also use the natbib \citep
%% and \citet commands to identify citations.  The citations are
%% tied to the reference list via symbolic KEYs. The KEY corresponds
%% to the KEY in the \bibitem in the reference list below. 

\section{Introduction} \label{sec:intro}
% Nomenclature: branches (entire 4.6um fundamental region), band structure/sub-bands (what we resolve), line (individual transitions).

JWST observations of exoplanet atmospheres have started to reveal absorption by carbon- and oxygen-bearing species at unprecedented precision. Recent measurements of the Saturn-mass hot Jupiter WASP-39b \citep{Faedi2011} showed the definitive detection of carbon dioxide (CO$_2$) \citep{JTEC_ERS_2022,alderson2022ers,rustamkulov2022ers} and water (H$_2$O) \citep{feinstein:niriss_ers,ahrer:nircam_ers,rustamkulov2022ers,alderson2022ers}, with additional absorption from photochemically generated SO$_2$  \citep{alderson2022ers,rustamkulov2022ers,tsai:2022}. The low-resolution, $R \,{\sim} 100$, NIRSpec PRISM observations additionally hinted at the presence of broadband absorption due to carbon monoxide (CO) \citep{rustamkulov2022ers}. However, the broad absorption structure of CO was unable to be robustly confirmed by the initial evaluation of the higher resolution NIRSpec G395H data, when binned to $R \,{\sim} 600$, likely due to blending of absorption from CO$_2$, H$_2$O, cloud opacity, and other high resolution molecular lines \citep{alderson2022ers}.

In hot dense atmospheres, like those of hot Jupiters, carbon is distributed between CO and CO$_2$ if C/O $\lesssim$ 1, with CO expected to be the dominant carbon-bearing molecule \citep{loddersfegley2002, heng2016carbon}. CO has been detected in the atmospheres of sub-stellar objects at high resolution from the ground for both transiting and non-transiting exoplanets \citep[e.g.,][]{snellen2010,brogi2014,dekok2013,Giacobbe2021,Line2021}, via direct imaging using a series of different methods \citep[e.g.,][]{konopacky2013,Snellen2015,barman2015, schwarz2015evidence,petitddlr2018,Hoeijmakers2018}, and in brown dwarf atmospheres \citep[e.g.,][]{cushing2005}. CO has also been seen in the upper atmospheres of all the Solar System's giant planets \citep{loddersfegley199book, encrenaz2004first}. More recently JWST spectra of the directly imaged and relatively isolated giant exoplanet VHS\,1256\,b \citep{miles2022} revealed the characteristic CO banding structure at 2.3 and 4.6 \textmu m using JWST's NIRSpec IFU, showing that it is possible to detect CO at these wavelengths and resolutions. 

Here we present a novel method to detect the fundamental ($\nu$=1-0) band structure of carbon monoxide at 4.6 \textmu m in the transmission spectrum of the hot Jupiter WASP-39b. Throughout this work we refer to the clusters of individual lines that make up the ro-vibrational band structure as sub-bands.

\section{JWST observations} \label{sec:observations}

Multiple transits of WASP-39b were observed by JWST as part of the Director’s Discretionary Early Release Science (ERS) program (ERS-1366; PIs: N. M. Batalha, J. Bean, and K. B. Stevenson) in July 2022 \citep{stevenson2016,bean2018}. Of these observations, a single transit was observed using JWST’s Near Infrared Spectrograph (NIRSpec) with the G395H grating \citep{jakobsen2022near, birkmann2022near}. This mode captures time-series spectra with a resolving power of $R \,{\sim} 2700$ (or ${\sim}6.6$\,\AA{ngstroms}/pixel) and a wavelength range from 2.75 to 5.16 \textmu m, making this observation well-suited to analysing the transmission spectrum of CO in higher resolution than ever before from space. This dataset is comprised of 465 integrations, each made up of 70 groups, and sums to a total exposure duration of 8.3 hours. This setup provides full coverage of the 2.8 hour transit event, as well as ample monitoring of the star's baseline flux pre- and post-transit.

NIRSpec's G395H grating disperses light across two adjacent detectors, NRS1 and NRS2, with a small ${\sim}0.1$\textmu m gap at $3.8$\textmu m. In search of CO signals, we focus our analysis on data longward of $3.8$\textmu m, that is, using data solely from the NRS2 detector. We reduce these data using a combination of the JWST Science Calibration Pipeline \citep[v1.6.2,][]{bushouse_howard_2022_7041998} and custom, open-source routines \citep[v0.1-beta,][]{lili_alderson_2022_7185855}. Following the JWST pipeline conventions, the data reduction, from raw data files to stellar spectra, is separated into two stages\footnote{Stage 1: processes 4D uncalibrated data into 3D rate-images.}\footnote{Stage 2: processes 3D rate-images into 2D stellar spectra.}, and is followed by light curve fitting.

\begin{figure}
\includegraphics[width=\columnwidth]{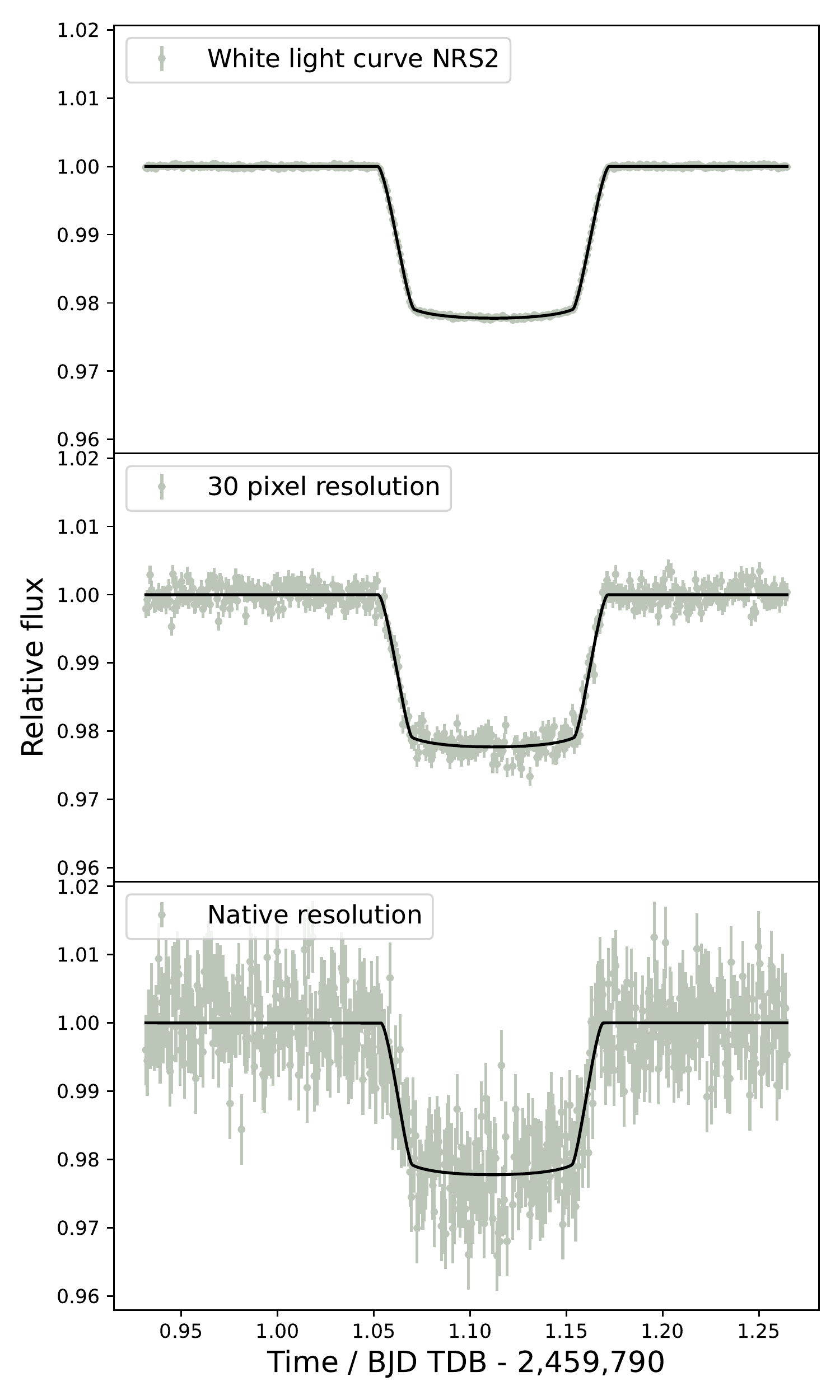}
\caption{Fits to the white light curve as well as example fits, centred on $4.7$\textmu m, to spectroscopic light curves at 30 pixel and native resolution. Shown are the corrected fluxes after the systematics model has been removed.}
\label{fig:light_curve}
\end{figure}

In stage 1 we largely utilise the JWST pipeline's default processing steps. However, we skip the dark-current step owing to the present quality of available reference files. Additionally, after the jump-detection step is run, we ``destripe'' the data at the group level. Destriping involves subtracting the column-by-column median background value from each column's pixels. This aims to reduce the 1/f noise at timescales longer than the column read time, or $440$\textmu s, and is more effective when corrected for at the group-level \citep[][]{JTEC_ERS_2022, rustamkulov2022ers}. These median values are computed with pixels having been flagged for poor data quality removed; and in particular, flagged pixels are made maximally consistent across groups within an integration. This flagging procedure ensures the same set of pixels are masked within a given pixel's ramp, and we find this improves ramp continuity in the presence of transient outliers such as cosmic ray hits. After destriping, the default ramp-fitting step is run and rate-images are output from stage 1.

In stage 2 we utilise custom routines to clean the rate-images and extract a time series of stellar spectra. Outlier cleaning is performed by finding deviations from an estimated spatial profile, as is common in optimal extraction \citep{horne1986optimal}. However, the curved spectral trace means the spatial profile must be created piecewise from windowed row-wise polynomial fits, as has been similarly utilised by \citet{zhang2021no}. We find a window size of 100 pixels, a polynomial order of 4, and an outlier threshold set at 4 standard deviations are sufficient to clean the curved spectral trace. Next, the background is subtracted column-by-column and the outlier cleaning is repeated once more. Stellar spectra are extracted using a box aperture centred on the spectral trace and with a width of 6 times the measured standard deviation of the point spread function. The resulting stellar spectra are comprised of between ${\sim}90,000$ and ${\sim}20,000$ electrons per column, decreasing from the short- to long-wavelength side of the NRS2 detector.

\subsection{Transmission spectrum} \label{subsec:planet_spectrum}

From the stellar spectra, we generate a transmission spectrum at the highest resolution possible, to maximise our chances of detecting the fine-structure of the CO absorption band. To achieve this, we make spectroscopic light curves at the native resolution of NIRSpec's G395H mode; these bins are just 2 pixels wide. For comparison, we also fit spectroscopic light curves with bins 30 pixels wide, as well as fit the entire NRS2 white light curve (see Figure \ref{fig:light_curve}). The light curves and transmission spectrum presented in \citet{alderson2022ers} consisted of bins 10 pixels wide. 

The data quality is such that systematics are barely visible in the raw light curves, with the exception of a small exponential ramp over the first ${\sim}10$ integrations, and a mirror tilt event at integration 269, as seen by \citet{alderson2022ers}. The exponential ramp is removed by discarding the first 15 integrations to ensure the full ramp effect does not impact the measured stellar baseline. The tilt event is accounted for by normalising the pre- and post-tilt flux by the pre- and post-transit median flux, respectively. We also discard three integrations immediately surrounding the tilt event. Any remaining systematics are modelled by linear regression of the measured relative positions of the spectral trace. These positions are found by cross-correlation in the dispersion direction using the median stellar spectra, and in the cross-dispersion direction using the median point spread function.

The total light curve model, $f(t)$, is described by
\begin{equation}
f(t) = T(t, \mathbf{\theta}) + \sum_{j \in W} f_{0, j} + S_j(x_j, y_j),
	\label{eq:light_curve_model_total}
\end{equation}
where
\begin{equation}
S_j(x_j, y_j) = p_{1, j} x_j + p_{2, j} y_j.
	\label{eq:light_curve_systematics}
\end{equation}
Here, $T$ is the physical transit model as a function of time, $t$, and transit parameter vector, $\mathbf{\theta}$. $f_0$ is an offset to the baseline flux. $S_j$ is the systematics model as a linear function of the spectral trace positions, $x_j$ and $y_j$. We elect to fit the non-physical model components piecewise, either side of the tilt event, where each window of data is denoted by the subscript $j$ from the set $W = \{\rm{pre{\text -}tilt}, \rm{post{\text -}tilt} \}$. The $x_j$ and $y_j$ traces can be seen in \citet[][extended data figures 1 and 2]{alderson2022ers}, and for computational efficiency the values are standardised by subtracting their means and dividing by their standard deviations, independently for each $j$. For $T$ we use a Batman transit model \citep{kreidberg2015batman} with a fixed non-linear 4-parameter limb-darkening law computed using ExoTiC-LD \citep{david_grant_2022_7437681} for each spectral bin using the 3D Stagger-grid of stellar models \citep{magic2015stagger}. 

\begin{figure}
\includegraphics[width=\columnwidth]{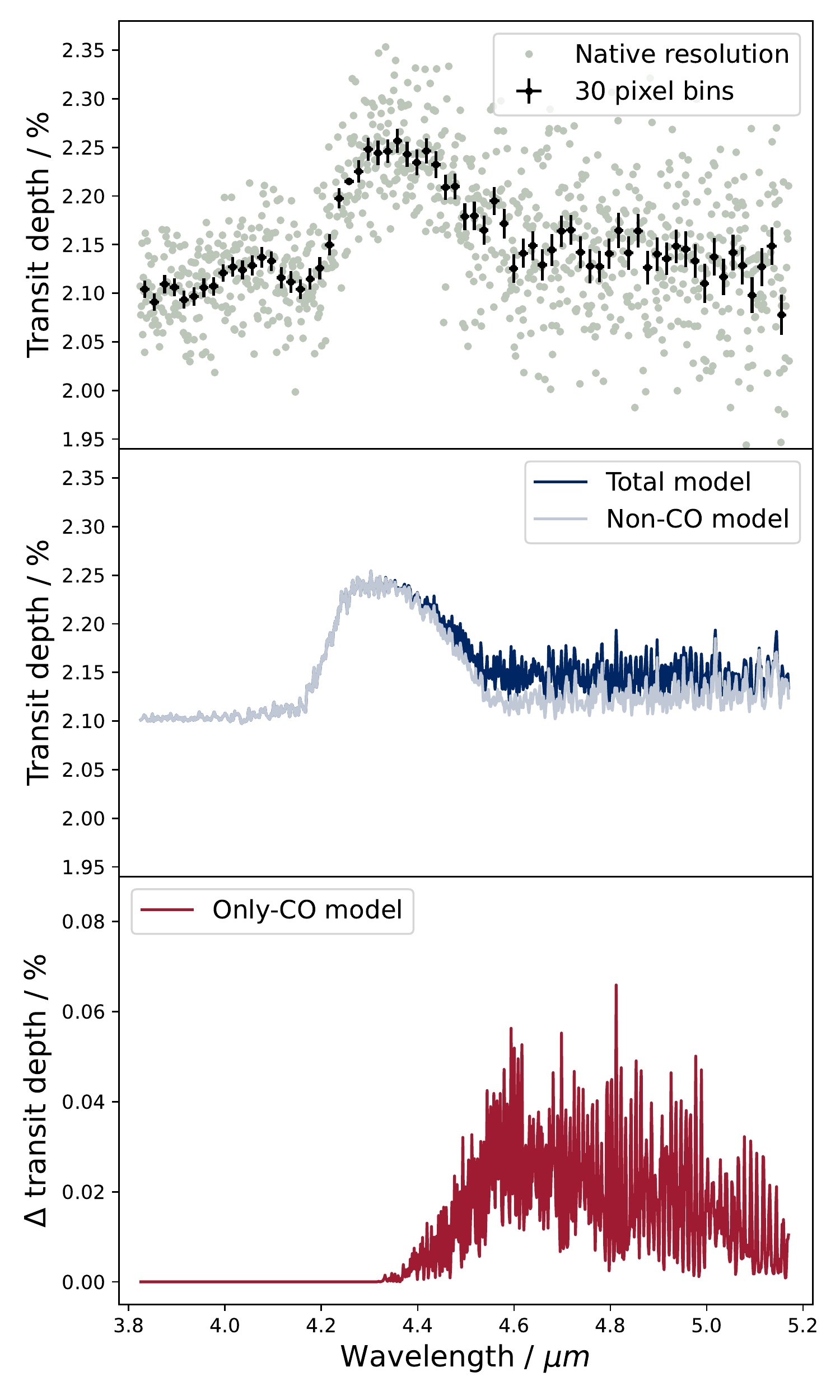}
\caption{Top panel: observed transmission spectra at native resolution (green) and at 30 pixel resolution (black). Middle panel: model transmission spectrum fit to the data (dark blue) and the same model but with CO removed (grey). Bottom panel: model contribution from CO only (red).}
\label{fig:transmission_spectra}
\end{figure}

\begin{figure*}
\centering
\includegraphics[width=1.9\columnwidth]{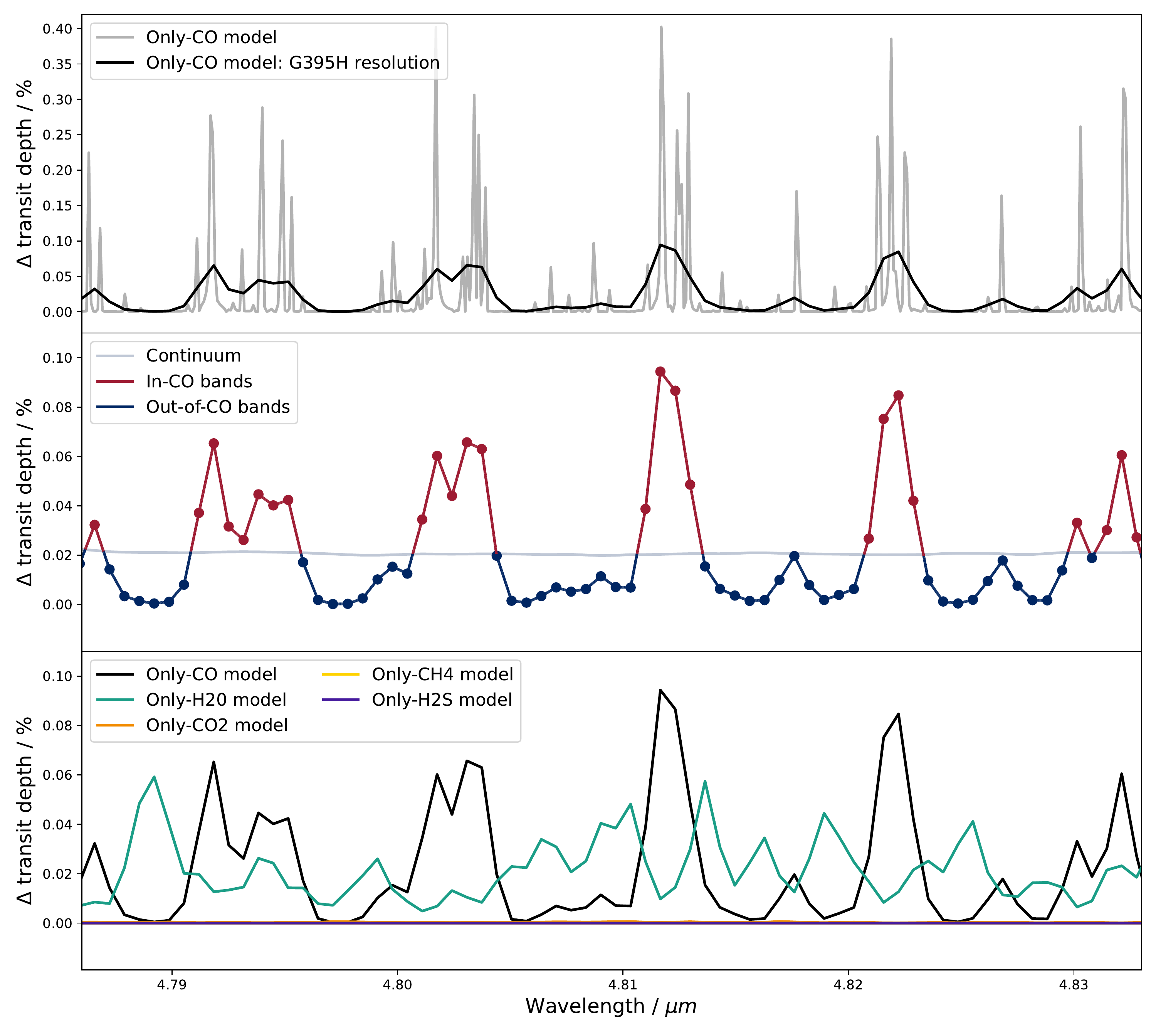}
\caption{Top panel: example wavelength sector showing the model contribution from CO only. The model is shown at simulated resolution (grey, sampled every \AA ngstrom) and after being convolved with the observed point spread function and rebinned onto the detector’s wavelength grid (black). Middle panel: visualisation of our method for detecting CO sub-band structure. Red and blue portions indicate wavelengths that are in-band or out-of-band, respectively. Note these wavelength selections are refined in Figure \ref{fig:co_lines}. Bottom panel: model contributions per molecule, where H$_2$O is the only significant molecular absorber in this wavelength sector other than CO.}
\label{fig:method_visualisation}
\end{figure*}

\begin{figure*}
\centering
\includegraphics[width=1.9\columnwidth]{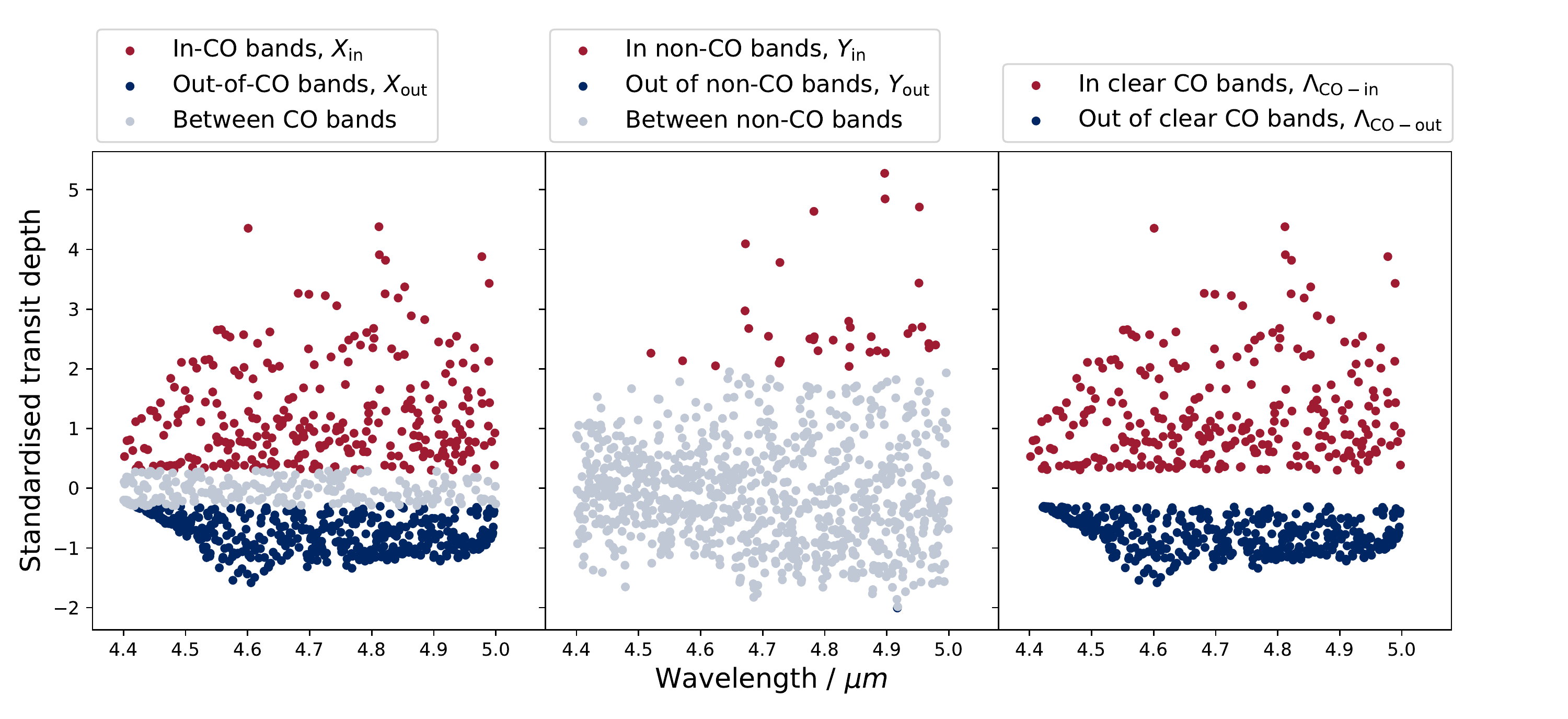}
\caption{Graphical representation of the logic behind our in-band (red) and out-of-band (dark blue) definitions. The panels show the cuts in the models for the only-CO model (left-hand panel), the non-CO model (middle panel), and the resulting clear CO sub-bands model (right-hand panel).
\label{fig:co_lines}}
\end{figure*}

We perform fitting of our light curve model with a least-squares optimiser \citep{2020SciPy-NMeth}, using the Levenberg-Marquadt algorithm \citep{more1978levenberg}, co-fitting the parameters $\mathbf{\theta}$, $f_{0, j}$, $p_{1, j}$, and $p_{2, j}$ simultaneously. We initially fit the white light curve to find the best-fit values for the center of transit time, $t_0 = 2459791.112128 \pm 0.000021 \,\mathrm{BJD}_\mathrm{TDB}$, semi-major axis, $a = 11.395 \pm 0.028 \,\mathrm{R_*}$, and inclination $i = 87.759 \pm 0.028^{\circ}$. These values are then fixed in the spectroscopic light curve fits that follow. For each spectroscopic light curve fit, the standard deviation of the residuals is used to inflate the original uncertainties on the fluxes, and then the fits are repeated. The uncertainties in the resulting transit depths are computed from the square root of the inverse Hessian matrix. We validate this fast approximation by using a Markov chain Monte Carlo algorithm \citep{foreman2013emcee}, with uniform priors, to sample the posterior distributions of several transit depths, and we find that the results are equivalent. These checks span the entire wavelength range of NRS2, with the inferred transit depths being on average 0.03 standard deviations discrepant and the corresponding uncertainties 1.7\% different.

In Figure \ref{fig:light_curve} we show fits to the white light curve, as well as example fits, centred on $4.7$\textmu m, to spectroscopic light curves at 30 pixel and native resolution. In each panel, we show the corrected flux values with the systematics model removed. The systematics model accounts for on average $1.3 \%$ of the residual variance, although close inspection of the light curves indicates a small amount of correlated noise remains. Tests varying the width of the spectral bins reveal the variance in the light curves becomes closer to the expected photon noise for narrower bins. This is indicative of the remaining noise being spatially correlated on the detector, most likely due to a small amount of leftover uncleaned 1/f noise. There is also likely some minor time-correlated noise, most clearly seen at the 30 pixel resolution, however we were unable to identify the origin of this noise. The mean residual standard deviations are 163 ppm, 1260 ppm, and 4644 ppm for the white light, 30 pixel, and native resolution light curve fits, respectively. At native resolution this precision corresponds to 1.1x photon noise. In the top panel of Figure \ref{fig:transmission_spectra} we show the transmission spectrum. 

\section{Theoretical models} \label{sec:models}

To investigate the presence of CO in the transmission spectrum of WASP-39b, we fit the observations to a grid of PHOENIX atmosphere models \citep{hauschildt:1999, barman:2001, lothringer:2020}. This was the same grid used in the initial investigation of WASP-39b's G395H spectrum in \cite{alderson2022ers}. The grid is sampled every \AA{ngstrom} and explores the irradiation temperature (i.e., absorbed heat from the star, $T_{\rm{irr}}$ = 1020, 1120, and 1220~K, respectively), internal temperature ($T_{\rm{int}}$ = 200 and 400~K), metallicity (0.1x, 1x, 10x, 50x, and 100x solar \citep[][]{asplund:2009}), C/O ratio (0.3, 0.54 (solar), 0.7, and 1.0), the presence of aerosols (a cloud deck at 0.3, 1, 3, and 10 mbar or a Rayleigh-scattering haze at 10x the H$_2$ scattering cross-section). We also included models with rainout chemistry (i.e., abundances of refractory elements are depleted in the layers above which they condense) compared to pure chemical equilibrium. Disequilibrium abundances of SO$_2$ from photochemistry \citep[][]{tsai:2022} were not included in these models. Species in the opacity calculation include CH, CH$_4$, CN, CO, CO$_2$, C$_2$, C$_2$H$_2$, C$_2$H$_4$, C$_2$H$_6$, CaH, CrH, FeH, HCN, HCl, HF, HI, HDO, HO$_2$, H$_2$, H$_2$S, H$_2$O, H$_2$O$_2$, H3$^+$, MgH, NH, NH$_3$, NO, N$_2$, N$_2$O, OH, O$_2$, O$_3$, PH$_3$, SF$_6$, SiH, SiO, SiO$_2$, TiH, TiO, VO and atoms up to U.  Of relevance for this work, we use the high-temperature CO linelist from \cite{goorvitch:1994} that was used for the PHOENIX grid in \cite{alderson2022ers}, but we also test against the newer \cite{li:2015} list. 

Our best-fit model at low-resolution, which we take to be our fiducial model, has $f$=0.351, $T_{\rm{int}}$ = 200~K, 10x solar metallicity, C/O = 0.3, a cloud deck at 0.3 mbar, no hazes, and rainout chemistry. These results are mostly consistent with those found using the median reduced G395H spectrum from 11 different analyses in \cite{alderson2022ers}, except for the internal temperature (200 versus 400~K) and the preference for rainout chemistry, both of which have relatively minor effects in the wavelength region probed by these G395H observations. We attribute this non-statistically-significant difference to the minor differences between data reductions.

We show the best-fit model compared to 30-pixel wide bins of the observations in the middle panel of Figure~\ref{fig:transmission_spectra}. Absorption from SO$_2$ at 4.0 \textmu m and CO$_2$ at 4.3 \textmu m are seen in the data, as previously described in \cite{JTEC_ERS_2022,alderson2022ers,rustamkulov2022ers, tsai:2022}. The middle panel also shows the model with and without the CO opacity. The difference of these two models is plotted in the bottom panel, indicating the contribution of CO to the best-fit transit spectrum. At the low effective resolution of 30 pixel bins, the presence of CO in the G395H spectrum is not apparent and its detection was not statistically justified due to blending with non-CO opacities like H$_2$O, CO$_2$ and the cloud opacity \citep{alderson2022ers}. 
%Any opacities absorbing in this region but not included in the non-CO model, like OCS, can be a source of noise when isolating the CO absorption.

\section{Defining CO sub-bands} \label{sec:defining_lines}

In order to detect the high-resolution features of CO, we define a set of wavelengths that are theorized to either display CO absorption (hereafter in-band) or not to display CO absorption (hereafter out-of-band). The aim is to compare the transit depths from each set, and thereby probe CO's fundamental band structure. This method is displayed graphically in Figure \ref{fig:method_visualisation}. This plot shows how at the resolution of the G395 mode we are able to probe small clusters of CO lines, which together we refer to as sub-bands.

We set out to define these sub-bands such that they represent a good contrast between the in- versus out-of-band transit depth and, furthermore, are not biased towards finding other potential species' high-resolution signals included in the atmospheric models. As we are concerned only with the high-resolution structure, we consider these sub-bands relative to the continuum level. We also restrict our analysis to the wavelength interval $4.4$\textmu m $ < \lambda < 5.0$\textmu m as this is the region that optimises the balance between the observed signal-to-noise and the amplitude of the model's high-resolution CO structure. In this region the resolving power of the G395 mode ranges from $R \,{\sim} 3040$ to ${\sim} 3516$ with increasing wavelength.

Now, with the models for only-CO and non-CO in hand from Section \ref{sec:models}, we layout a method for defining clear CO sub-bands. For each model, we remove the continuum with a Savitzky-Golay filter \citep[][]{Savitzky1964SmoothingProcedures} and then standardise the mean and standard deviation of the model transit depths. To define the clear in-band wavelengths, $\Lambda_{\rm{CO{\text -}in}}$, we select the only-CO model transit depths above a set threshold, but only where these wavelengths do not show appreciable signals in the non-CO model. To define the clear out-of-band wavelengths,  $\Lambda_{\rm{CO{\text -}out}}$, we use the same logic, but for only-CO model points below a set threshold. In this way, the sub-bands will probe the highest contrast regions of the CO signal, without inadvertently confusing the CO signal with other species. In terms of logical operators we can write
\begin{eqnarray}
\Lambda_{\rm{CO{\text -}in}} &&= X_{\rm{in}} \,\rm{\textbf{\scalebox{0.9}{AND}}}\, (\,\rm{\textbf{\scalebox{0.9}{NOT}}}\, (Y_{\rm{in}} \,\rm{\textbf{\scalebox{0.9}{OR}}}\, Y_{\rm{out}})),\\
\label{eq:line_definition_in_co}
\Lambda_{\rm{CO{\text -}out}} &&= X_{\rm{out}} \,\rm{\textbf{\scalebox{0.9}{AND}}}\, (\,\rm{\textbf{\scalebox{0.9}{NOT}}}\, (Y_{\rm{in}} \,\rm{\textbf{\scalebox{0.9}{OR}}}\, Y_{\rm{out}})),
\label{eq:line_definition_out_co}
\end{eqnarray}
where $X_{\rm{in}}$ and $X_{\rm{out}}$ are wavelengths with only-CO model transit depths above and below set thresholds, and $Y_{\rm{in}}$ and $Y_{\rm{out}}$ are wavelengths with non-CO model transit depths above and below set thresholds, respectively. 

As a graphical description, we display the selected wavelengths in Figure \ref{fig:co_lines}. Here, the clear in- and out-of-band wavelengths are shown in the right-hand panel. We set the threshold values to $\pm 0.3$ for $X$ and to $\pm 2.0$ for $Y$. The motivation for these threshold values is to balance the contrast of transit depth between in-band and out-of-band versus the number of bands included in the samples, and thus the statistical power of any detection test. These thresholds result in 259 clear bands, of which 111 are in-band and 148 are out-of-band. The distribution of the number of contiguous pixels in-band has a mean of 3.0 pixels and a standard deviation of 1.1 pixels. The distribution for out-of-band has a mean of 3.8 pixels and a standard deviation of 1.7 pixels.

\section{Detecting CO sub-bands} \label{sec:detection}

\begin{figure}
\centering
\includegraphics[width=\columnwidth]{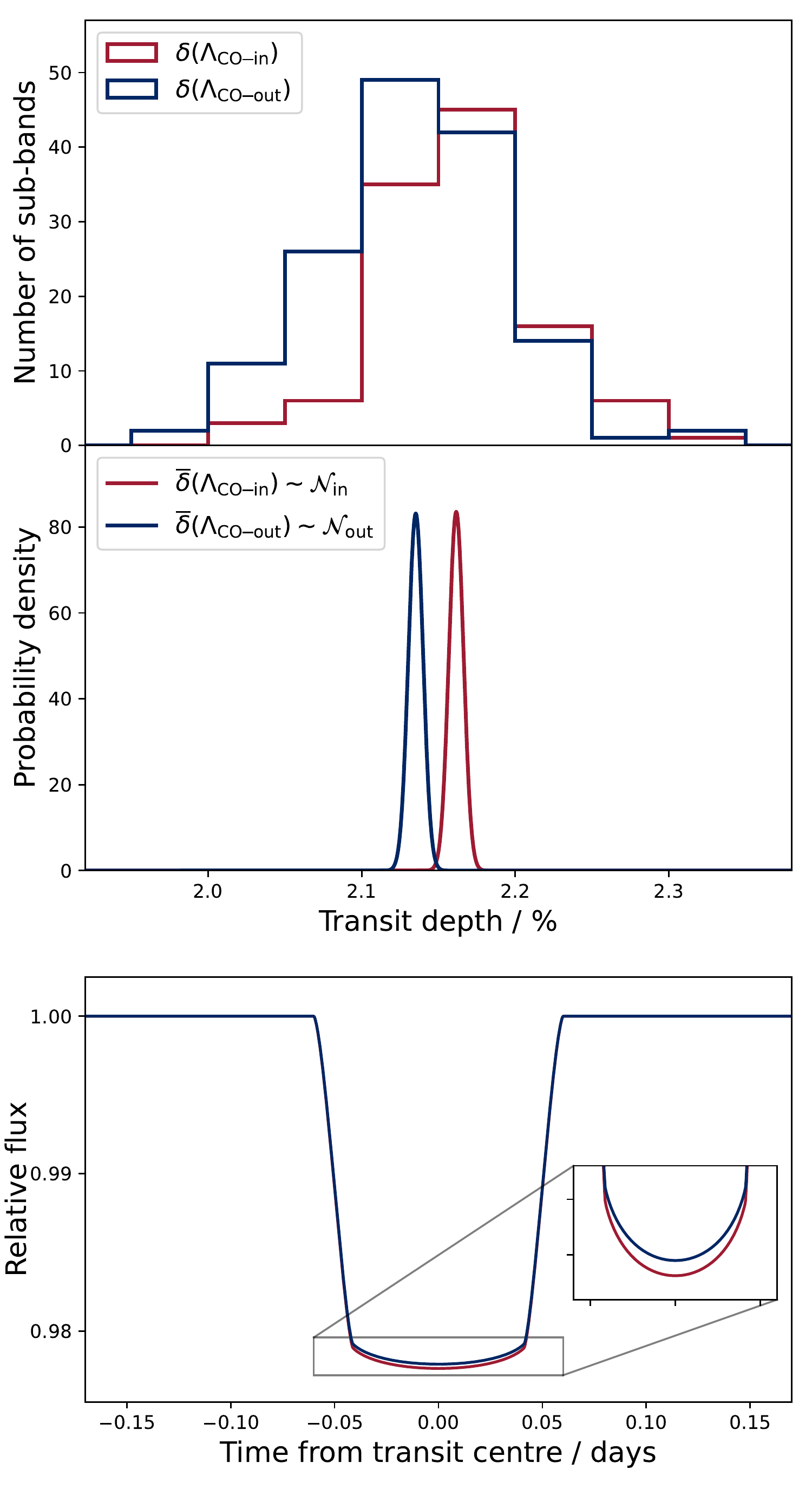}
\caption{Top panel: samples showing the number of sub-bands as a function of transit depth, $\delta$, for in- and out-of-band CO. Middle panel: sampling distributions of the mean transit depths, $\overline{\delta}$, for in- and out-of-band CO. These sampling distributions are modelled as Gaussians with means and standard deviations inferred from the measured samples shown in the top panel. Bottom panel: light curve models visualising the measured difference in transit depth for in- and out-of-band CO.}
\label{fig:co_detection}
\end{figure}

Using our definition of CO sub-bands, we categorize the transit depths from the observed native-resolution transmission spectrum as either in-band or out-of-band. This process generates two samples of transit depths, one for each of the theorized underlying populations of transit depth for in-band and out-of-band. As in Section \ref{sec:defining_lines}, we remove the continuum from the observed transit depths to probe only the high-resolution CO features. We display these two samples in the top panel of Figure \ref{fig:co_detection}. The CO in-band sample, $\delta(\Lambda_{\rm{CO{\text -}in}})$, has a mean, $\mu_{\rm{in}}=21616$ ppm, and standard deviation, $\sigma_{\rm{in}}=501$ ppm. The CO out-of-band sample, $\delta(\Lambda_{\rm{CO{\text -}out}})$, has a mean, $\mu_{\rm{out}}=21352$ ppm, and standard deviation, $\sigma_{\rm{out}}=576$ ppm.

To test statistically if the in-band sample displays greater CO absorption than the out-of-band sample, we conduct a hypothesis test. We run a two-sample T-test, specifically a one-tailed Welch's t-test \citep{welch1947generalization}, with the following hypothesis:
\begin{eqnarray}
\mathcal{H}_0 &&: \mu_{\rm{in}} = \mu_{\rm{out}},\\
\label{eq:null_hypothesis}
\mathcal{H}_1 &&: \mu_{\rm{in}} > \mu_{\rm{out}},
\label{eq:alternate_hypothesis}
\end{eqnarray}
where $\mathcal{H}_0$ and $\mathcal{H}_1$ are the null and alternate hypotheses, respectively. Running this test on our samples, we find a statistically significant result. The in-band samples have a greater mean transit depth than the out-of-band samples, with $t=3.90$ and $p=0.00006$, and hence we detect high-resolution CO band structure with assurance. 

As a visual representation, we plot the sampling distributions of the mean transit depths in the middle panel of Figure \ref{fig:co_detection}. The sampling distributions are normally distributed such that
\begin{eqnarray}
\overline{\delta}(\Lambda_{\rm{CO{\text -}in}}) && \sim \mathcal{N} \Big( \mu_{\rm{in}}, \frac{\sigma^2_{\rm{in}}}{n_{\rm{in}}} \Big),\\
\label{eq:sampling_dist_in_co}
\overline{\delta}(\Lambda_{\rm{CO{\text -}out}}) && \sim \mathcal{N} \Big( \mu_{\rm{out}}, \frac{\sigma^2_{\rm{out}}}{n_{\rm{out}}} \Big),
\label{eq:sampling_dist_out_co}
\end{eqnarray}
where the standard errors are $\sigma_{\rm{in}} / n_{\rm{in}}^{1/2} = 48$ ppm and $\sigma_{\rm{out}} / n_{\rm{out}}^{1/2}=48$ ppm, and the sample sizes are given by $n_{\rm{in}}$ and $n_{\rm{out}}$. The difference between the transit depth means is $264 \pm 68$ ppm. This difference, in terms of the resulting light curve models, is displayed in the bottom panel of Figure \ref{fig:co_detection}.

Further to this result, we conduct sensitivity tests for the $X$ and $Y$ thresholds set in Section \ref{sec:defining_lines}. We find that our results are not sensitive to the $Y$ threshold, but when varying the $X$ threshold we observe a change in the detected transit depth difference between in-band and out-of-band samples. As the $X$ threshold increases the transit depth difference increases, but with increasing uncertainty owing to the reduced statistical power. For example, if the $X$ threshold values are increased to $\pm 0.4$, the difference between the transit depth means becomes $301 \pm 81$ ppm. Intuitively this effect can be understood as the larger thresholds narrowing in on the peaks of the CO sub-bands, where the contrast in transit depth is greater but fewer data points lie.

\section{Discussion} \label{sec:discussion}

The technique presented works particularly well with JWST's NIRSpec G395H mode, owing to its comparatively high spectral resolution. Given this resolution, it was vital in our analysis to account for the Doppler transformations of both the barycentric velocity of the WASP-39 system, $\gamma_* = -58.4421 \,\rm{\,km \, s^{-1}}$ \citep{mancini2018gaps}, and of JWST's barycentric velocity at the time of observation, $\gamma_{\rm{\scalebox{0.6}{JWST}}} = -28.88807 \,\rm{\,km \, s^{-1}}$. In fact,
the robustness of our detection is further backed up by trialing spurious Doppler transformations. We find that shifting our models by velocities equivalent to just one resolution element ($\pm 88 \,\rm{\,km \, s^{-1}}$ at 4.5 \textmu m) in either direction leads to non-detections, with $p=0.42$ and $0.17$. This further highlights the fidelity and soundness of our detection, whilst also underscoring the need for models applied to the highest resolution JWST modes to be appropriately Doppler shifted when searching for narrow molecular features.

Our technique offers an independent test for the presence of molecular features in the atmospheres of exoplanets for well characterised molecules with precise line lists. Differences in the measured excess transit depth of only 4 ppm were found using the different CO line lists of \cite{goorvitch:1994} and \cite{li:2015}, though such differences may be greater in the overtone bands where the two lists disagree more than in the fundamental band. The technique may be applied to other molecules with rotational sub-band structures distinctly separated at $R \,{\sim} 3000$, for example CH$_4$, NH$_3$, or HCN between 2.5 and 3.5 \textmu m, but the development and use of accurate and complete line lists for other species is necessary. The structure may be more difficult to measure for molecules such as H$_2$O that are more easily distinguished by their broadband structure due to the ro-vibrational lines being more tightly clustered. 

Aside from the technique outlined in this work, cross-correlation methods, often used with high-resolution ground-based data, may also be applied to the highest resolution JWST modes. Certain cross-correlation formalisms \citep[e.g.,][]{brogi2019retrieving, gibson2020detection} enable the construction of a likelihood function which can also be used for the detection and retrieval of molecules and their abundances, and a future methodology comparison with traditional low-resolution methods may be fruitful.

\section{Summary and conclusions} \label{sec:conclusion}

We have detected absorption structure in CO's fundamental band using JWST's NIRSpec G395H mode. To make this detection we employed a new technique in which we categorized wavelength bins as either in-band or out-of-band, and then compared the transit depths of the resulting samples. We found the mean transit depth of CO sub-bands to be $264 \pm 68$ ppm larger than the transit depth between CO sub-bands. This value is in agreement with the prediction from the PHOENIX model spectrum of $217$ ppm, when using the same sub-bands.

While CO$_2$'s opacity is much stronger per molecule in the G395H wavelength regime and was therefore more easily detected in \cite{alderson2022ers}, CO is actually by-far the dominant carbon-bearing molecule throughout WASP-39b's atmosphere with a 1 mbar volume mixing ratio in the best-fit model of 2,326 ppm compared to 7 ppm of CO$_2$. This fact remains true for all giant planets in chemical equilibrium with plausible heavy-element enrichments (e.g., $M/H \leq 100$x solar). Thus the detection of CO, in addition to CO$_2$, enables a more robust and complete understanding of a planet's chemical inventory, especially with respect to the measurement of the C/O ratio. Without such a measurement, any inference of the bulk carbon abundance or C/O ratio from CO$_2$ alone would rely on assumptions of chemical equilibrium. We can therefore not only be more confident in the detection of CO with JWST/NIRSpec/PRISM \citep[][]{rustamkulov2022ers}, but also in the conclusions from the other JWST spectra that WASP-39b has a sub-solar to solar C/O ratio and likely formed interior to the CO$_2$ ice line \citep[][]{alderson2022ers,ahrer:nircam_ers,feinstein:niriss_ers}.

%% IMPORTANT! The old "\acknowledgment" command has be depreciated. It was
%% not robust enough to handle our new dual anonymous review requirements and
%% thus been replaced with the acknowledgment environment. If you try to 
%% compile with \acknowledgment you will get an error print to the screen
%% and in the compiled pdf.
%% 
%% Also note that the akcnowlodgment environment does not support long amounts of text. If you have a lot of people and institutions to acknowledge, do not use this command. Instead, create a new \section{Acknowledgments}.
\section*{Acknowledgments}
We thank the two anonymous reviewers for helpful comments. We thank the JWST Transiting Exoplanet Community ERS team for their hard work on the proposal, observations, and data presented in this paper. We thank the JTEC ERS G395H sub-team for fruitful conversations. This work is based on observations made with the NASA/ESA/CSA JWST. The data were obtained from the Mikulski Archive for Space Telescopes at the Space Telescope Science Institute, which is operated by the Association of Universities for Research in Astronomy, Inc., under NASA contract NAS 5-03127 for JWST. These observations are associated with program JWST-ERS-01366. Support for program JWST-ERS-01366 was provided by NASA through a grant from the Space Telescope Science Institute. The specific observations analysed can be accessed via \dataset[DOI: 10.17909/1j77-6w13]{https://doi.org/10.17909/1j77-6w13}. Science data processing version (SDP\_VER) 2022\_2a generated the uncalibrated data that we downloaded from MAST. We used JWST Calibration Pipeline software version (CAL\_VER) 1.6.2 with modifications described in the text. We used calibration reference data from context (CRDS\_CTX) 0930, except as noted in the text. D. Grant acknowledges funding from the UKRI STFC Consolidated Grant ST/V000454/1. The data and models associated with this study can be found in The JWST Transiting Exoplanet Community Early Release Science Program Zenodo Library. 

Author Contributions: DG performed the data analysis with atmospheric models supplied by JDL. DG, JDL and HRW wrote the manuscript. DG and JDL designed the analysis method in discussion with the ERS G395H sub-team led by HRW and LA. All other authors read and approved the manuscript. NMB, JLB, and KBS provided overall program leadership of the JWST Transiting Exoplanet Community ERS program.

%% To help institutions obtain information on the effectiveness of their 
%% telescopes the AAS Journals has created a group of keywords for telescope 
%% facilities.
%
%% Following the acknowledgments section, use the following syntax and the
%% \facility{} or \facilities{} macros to list the keywords of facilities used 
%% in the research for the paper.  Each keyword is check against the master 
%% list during copy editing.  Individual instruments can be provided in 
%% parentheses, after the keyword, but they are not verified.

\vspace{5mm}
\facility{JWST(NIRSpec G395H).}

%% Similar to \facility{}, there is the optional \software command to allow 
%% authors a place to specify which programs were used during the creation of 
%% the manuscript. Authors should list each code and include either a
%% citation or url to the code inside ()s when available.

\software{JWST pipeline \citep{bushouse_howard_2022_7041998}, ExoTiC-JEDI \citep[][]{lili_alderson_2022_7185855}, ExoTiC-LD \citep{david_grant_2022_7437681}, Batman \citep{kreidberg2015batman}, numpy \citep[][]{harris2020array}, SciPy \citep[][]{2020SciPy-NMeth}, matplotlib \citep[][]{Hunter:2007}, pandas \citep{jeff_reback_2022_6702671}, xarray \citep{hoyer2017xarray, hoyer_stephan_2022_6323468}, astropy \citep{2013A&A...558A..33A,2018AJ....156..123A, price2022astropy}.}

\bibliography{sample631}{}
\bibliographystyle{aasjournal}

%% This command is needed to show the entire author+affiliation list when
%% the collaboration and author truncation commands are used.  It has to
%% go at the end of the manuscript.
%\allauthors

%% Include this line if you are using the \added, \replaced, \deleted
%% commands to see a summary list of all changes at the end of the article.
%\listofchanges

\end{document}